\documentclass{rstransa}

\begin{document}

\title{Development of research network on Quantum Annealing Computation and Information using Google Scholar data}

\author{
Antika Sinha$^{1}$}

\address{$^{1}$Department of Computer Science, Asutosh College, Kolkata 700026, India}
\subject{statistics, complexity}


\corres{Antika Sinha\\
\email{antikasinha@gmail.com}}

\begin{abstract}
 We build and analyze the network of hundred top cited nodes (research  papers and books from Google Scholar; strength or citation of the nodes range from about 44000 up to 100) starting early 1980 to till last year. These searched publications (papers, books) are based on Quantum Annealing Computation and Information categorized in four different sets: A) Quantum/Transverse Field Spin Glass Model, B) Quantum Annealing, C) Quantum Adiabatic Computation and  D) Quantum Computation Inform-\\ation in the title or abstract of the searched publications. We fitted the growth in the annual number of publication ($n_p$) in each of these four categories A to D to the form $n_p \sim\exp{(t/\tau)}$ where $t$ denotes the time in year. We found the scaling time $\tau$ to be of order about 10 years for category A and C whereas $\tau$ is order of about 5 years for category B and D.
\end{abstract}

\begin{fmtext}

\section{Introduction}

Though the classical gate-based digital computation
has been extremely successful, by early eighties,
serious limitations of the classical gate-based
computers started to became obvious for the so-called
(computationally) hard problems. Some of the prominent
examples being that of $N$-city Traveling Salesman
Problem (of searching the minimum travel distance
path from $N!$ order candidates) or the ground
states in randomly frustrated $N$ Ising-spin glasses
(search for minimum energy state out of the $2^N$
states) where search time is not bounded by a polynomial in
$N$. 

Inspired by the metallurgical annealing technique,
Kirkpatrick, Gelatt and Vecchi~\cite{1} in 1983 proposed
a novel stochastic or statistical physical technique,
post which researchers attempted to achieve stochastic 

\end{fmtext}

\maketitle

\noindent
practical computational solutions of such hard
multi-variable optimization problems.  Though some
practical solutions could be obtained using this technique very quickly,  the search  
time still could not be
bounded by any polynomial (in $N$).

 Parallelly, Feynman~\cite{2} in 1982
proposed quantum gate based computing
architectures for solving the computationally 
hard problems. So far, such gate based quantum
computers are not very implementable or achieved
any practical level. In 1989, Ray, Chakrabarti and
Chakrabarti~\cite{3} proposed
a quantum tunneling induced stochastic search
algorithm for the search of the ground state of
the Sherrington-Kirkpatrick spin glass model.
Although, there were many criticisms of this
idea, the group of Aeppli and Rosenbaum~\cite{4} in 1991  experimentally demonstrated
its feasibility. In 1998, Kadwaki
 and Nishimori~\cite{5} numerically showed
 the superiority  of the quantum
annealing technique over the classical annealing
technique and finally, Johnson et al,~\cite{6} developed and marketed the first quantum annealing machine (D-Wave computers)
in 2010. Since then a revolution in quantum
computing and information processing has taken
place.

\begin{figure}[!htbp]
 \includegraphics[width=14.5cm,height=6.5cm]{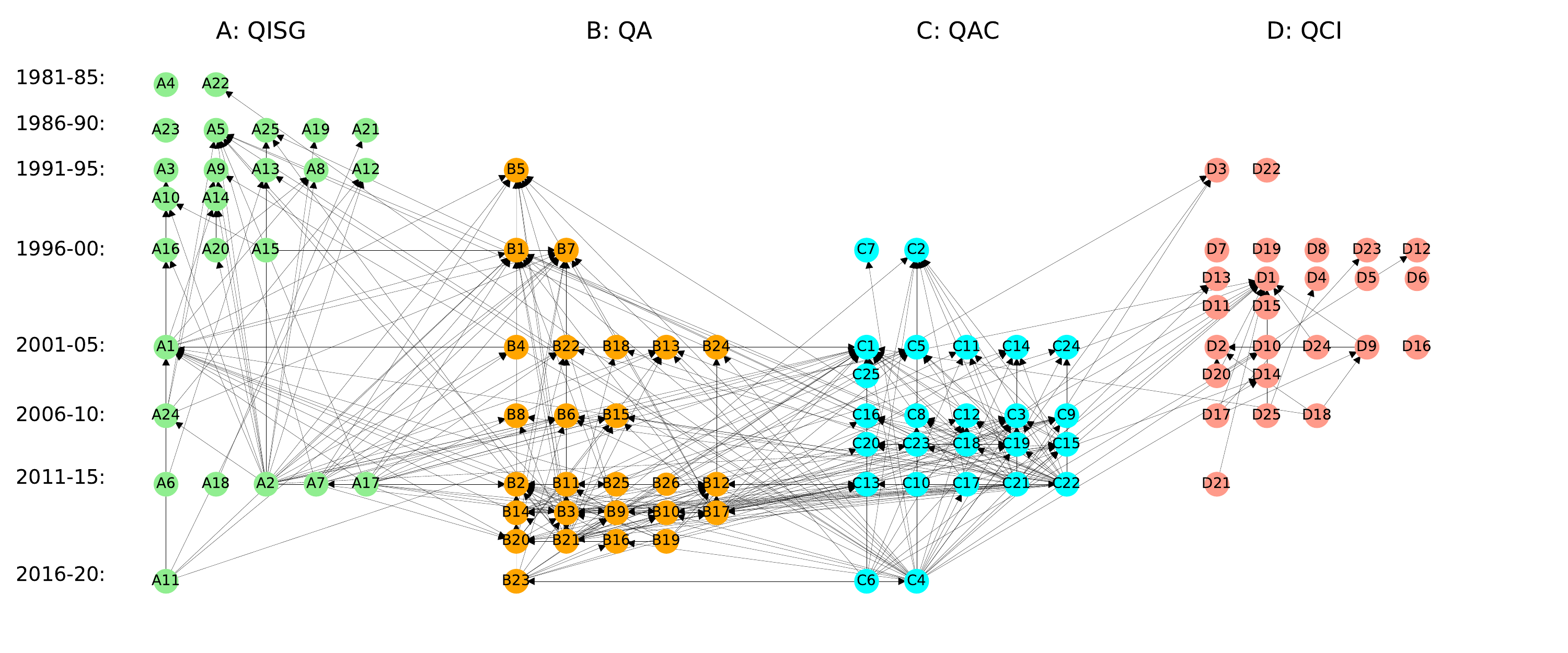}
  \caption{ 
  Network of hundred top cited (in the range of about 44000 to 100) nodes or research publications (papers, books) from Google scholar (date of access: 10 April-2022) on  ``Quantum Annealing Computation and Information'' from 1980 to 2020. These nodes are classified into four different categories (A to D): A. QISG (Quantum/Transverse Field Spin Glass Model), B. QA (Quantum Annealing), C. QAC (Quantum Adiabatic Computation) and D. QCI (Quantum Computation Information). These nodes for the last forty years are further subdivided into five years slot according to the respective years of publication. The network links (from citing to cited nodes) are also shown.  }
  \label{fig-timeline}
 \end{figure}

Here, we study the growth of research publications (papers, books) of the well studied problem-area ``Quantum Annealing Computation and Information'' from 1980 up to 2021 using annual Number of Publications ($n_p$) and a research network containing hundred top cited nodes as publications and the corresponding edges as citations.  
All data are collected between 8 to 10 April-2022, using the popular and publicly available search engine Google scholar, where the citation strength of the searched nodes are seen to vary nearly between 44000 to 100. The set of hundred top cited publications are categorized into four different categories (A to D) considering twenty five publications from each category as: A) Quantum/Transverse Field Spin Glass Model, B) Quantum Annealing, C) Quantum Adiabatic Computation and D) Quantum Computation Information. $n_p$ from each categories are observed to follow exponential distribution against time in years and best fits as $n_p \sim\exp{(t/\tau)}$. We find that the `similarity classes' of growth time constant (similar value of $\tau$) exists for category A and C and for category B and D that we discuss further in the paper.

\begin{table}[!htbp]
    \caption{Citation data of the nodes/research publications (papers, books) for Categories  A and B (date of data collection is 10 April, 2022).}
    \label{tab-tableAB}
    \begin{tabular}{|c|c|c|c||c|c|c|c|} 
    \hline
       \multicolumn{4}{|c||} {\textbf{Category A (QISG)}} & \multicolumn{4}{c|} {\textbf{Category B (QA)}} \\ \cline{1-8} \textbf{paper id} & \textbf{year} & \textbf{citation} & \textbf{reference} & \textbf{paper id} & \textbf{year} & \textbf{citation} & \textbf{reference} \\
      \hline
		A1 & 2002 & 641 & \cite{7} & B1 & 1998 & 1570 & \cite{5}  \\
		\hline     
		A2 & 2015 & 316 & \cite{8} & B2 & 2011 & 1431 & \cite{6} \\
		\hline
		A3 & 1991 & 256 & \cite{4} & B3 & 2014 & 711 & \cite{9} \\
		\hline
		A4 & 1982 & 213 & \cite{10} & B4 & 2002 & 641 & \cite{7} \\
		\hline
		A5 & 1989 & 207 & \cite{3} & B5 & 1994 & 630 & \cite{11} \\
		\hline
		A6 & 2011 & 199 & \cite{12} & B6 & 2008 & 594 & \cite{13}  \\
		\hline
		A7 & 2015 & 196 & \cite{14} & B7 & 1999 & 533 & \cite{15}  \\
		\hline
		A8 & 1994 & 185 & \cite{16} & B8 & 2006 & 333 & \cite{17} \\
		\hline
		A9 & 1993 & 179 & \cite{18} & B9 & 2014 & 327 & \cite{19} \\
		\hline
		A10 & 1995 & 177 & \cite{20} & B10 & 2014 & 309 & \cite{21} \\
		\hline
		A11 & 2018 & 174 & \cite{22} & B11 & 2012 & 305 & \cite{23}  \\
		\hline
		A12 & 1994 & 172 & \cite{24} & B12 & 2013 & 282 & \cite{25} \\
		\hline
		A13 & 1993 & 164 & \cite{26} & B13 & 2005 & 268 & \cite{27}  \\
		\hline
		A14 & 1995 & 161 & \cite{28} & B14 & 2013 & 227 & \cite{29}  \\
		\hline
		A15 & 2000 & 159 & \cite{30} & B15 & 2008 & 225 & \cite{31}  \\
		\hline
		A16 & 1996 & 154 & \cite{32} & B16 & 2015 & 202 & \cite{33}  \\
		\hline
		A17 & 2015 & 147 & \cite{34} & B17 & 2014 & 187 & \cite{35}  \\
		\hline
		A18 & 2014 & 138 & \cite{36}  & B18 & 2004 & 185 & \cite{37} \\
		\hline
		A19 & 1990 & 134 & \cite{38} & B19 & 2015 & 202 & \cite{39}  \\
		\hline
		A20 & 1996 & 107 & \cite{40} & B20 & 2014 & 169 & \cite{41} \\
		\hline
		A21 & 1990 & 105 & \cite{42} & B21 & 2014 & 169 & \cite{43}  \\
		\hline
		A22 & 1985 & 99 & \cite{44} & B22 & 2002 & 151 & \cite{45}  \\
		\hline
		A23 & 1987 & 93 & \cite{46} & B23 & 2017 & 149 & \cite{47} \\
		\hline
		A24 & 2008 & 91 & \cite{48} & B24 & 2005 & 126 & \cite{49}  \\
		\hline
		A25 & 1989 & 91 & \cite{50} & B25 & 2012 & 123 & \cite{51} \\
		\hline

		\end{tabular}
\end{table}

\begin{table}[!htbp]
    \caption{Citation data of the nodes/research publications (papers, books) for Categories  C and D (date of data collection is 10 April, 2022).}
    \label{tab-tableCD}
   \begin{tabular}{|c|c|c|c||c|c|c|c|} 
    \hline
    \multicolumn{4}{|c||} {\textbf{Category C (QAC)}} & \multicolumn{4}{c|}
    {\textbf{Category D (QCI)}} \\ \cline{1-8} \textbf{paper id} & \textbf{year} & \textbf{citation} & \textbf{reference} & \textbf{paper id} & \textbf{year} & \textbf{citation} & \textbf{reference} \\
    \hline
		C1 & 2001 & 1959 & \cite{52} & D1 & 2000 & 43720 & \cite{53}   \\
		\hline     
		C2 & 2000 & 1217 & \cite{54} & D2 & 2001 & 6569 & \cite{55} \\
		\hline
		C3 & 2008 & 739 & \cite{56} & D3 & 1992 & 3278 & \cite{57} \\
		\hline
		C4 & 2018 & 700 & \cite{58} & D4 & 2000 & 3232 & \cite{59} \\
		\hline
		C5 & 2001 & 517 & \cite{60} & D5 & 2000 & 2933 & \cite{61}  \\
		\hline
		C6 & 2016 & 413 & \cite{62} & D6 & 2000 & 2930 & \cite{63}  \\
		\hline
		C7 & 1998 & 408 & \cite{64} & D7 & 1996 & 1824 & \cite{65} \\
		\hline
		C8 & 2007 & 337 & \cite{66} & D8 & 1998 & 1620 & \cite{67} \\
		\hline
		C9 & 2008 & 326 & \cite{68} & D9 & 2003 & 1603 & \cite{69} \\
		\hline
		C10 & 2012 & 290 & \cite{70} & D10 & 2002 & 1415 & \cite{71} \\
		\hline
		C11 & 2001 & 266 & \cite{72} & D11 & 2000 & 1102 & \cite{73} \\
		\hline
		C12 & 2007 & 252 & \cite{74} & D12 & 1999 & 1089 & \cite{75} \\
		\hline
		C13 & 2011 & 242 & \cite{76} & D13 & 1999 & 1015 & \cite{77} \\
		\hline
		C14 & 2005 & 231 & \cite{78} & D14 & 2005 & 951 & \cite{79}  \\
		\hline
		C15 & 2010 & 218 & \cite{80} & D15 & 2000 & 911 & \cite{81} \\
		\hline
		C16 & 2006 & 169 & \cite{82} & D16 & 2003 & 837 & \cite{83} \\
		\hline
		C17 & 2014 & 169 & \cite{84} & D17 & 2007 & 774 & \cite{85} \\
		\hline
		C18 & 2009 & 167 & \cite{86} & D18 & 2009 & 761 & \cite{87} \\
		\hline
		C19 & 2009 & 162 & \cite{88} & D19 & 1997 & 685 & \cite{89}  \\
		\hline
		C20 & 2008 & 153 & \cite{90} & D20 & 2003 & 837 & \cite{91}  \\
		\hline
		C21 & 2014 & 151 & \cite{43} & D21 & 2015 & 663 & \cite{92}  \\
		\hline
		C22 & 2015 & 132 & \cite{93}  & D22 & 1995 & 650 & \cite{94} \\
		\hline
		C23 & 2008 & 129 & \cite{95} & D23 & 1998 & 645 & \cite{96} \\
		\hline
		C24 & 2005 & 112 & \cite{97} & D24 & 2002 & 580 & \cite{98} \\
		\hline
		C25 & 2005 & 110 & \cite{99} & D25 & 2007 & 99 & \cite{100} \\
		\hline

		\end{tabular}
\end{table}

\begin{table}[!htbp]
    \caption{Top ten nodes (publication) having more outdegree than indegree within the network.}
    \label{tab-tableinout}
   \begin{tabular}{|c|c|c|c|c|c|c|c|c|c|c|}
    \hline
        \multicolumn{11}{|c|} {\textbf{TOP 10 INDEGREE (Self-Category  Citation) }}   \\
    \hline
        paper id  & C1  & B2 & B1 & D1 & B7 & B5  & B3,~B12,~C3  & A1,~A5,~C2 & B14,~C15 & B15,~C5,~C9,~C13  \\
		\hline     
        indegree  & 22 & 16 & 15 & 14 & 12 & 11 & 10  &  9 & 8 & 7 \\
        \hline
        \multicolumn{11}{|c|} {\textbf{TOP 10 OUTDEGREE (Cross-Category Citation) }} \\ 
        \hline
        paper id  & C4 & C21 & B21 & C22 & A2 & B17  & A17 & B3 & B20  & B14   \\
		\hline     
        outdegree  & 29 & 28 & 27 & 26 & 25 & 20  & 19 & 14 & 13  &  12  \\
        \hline

		\end{tabular}
\end{table}

 \begin{figure}[!htbp]
 \includegraphics[width=8.5cm,height=6cm]{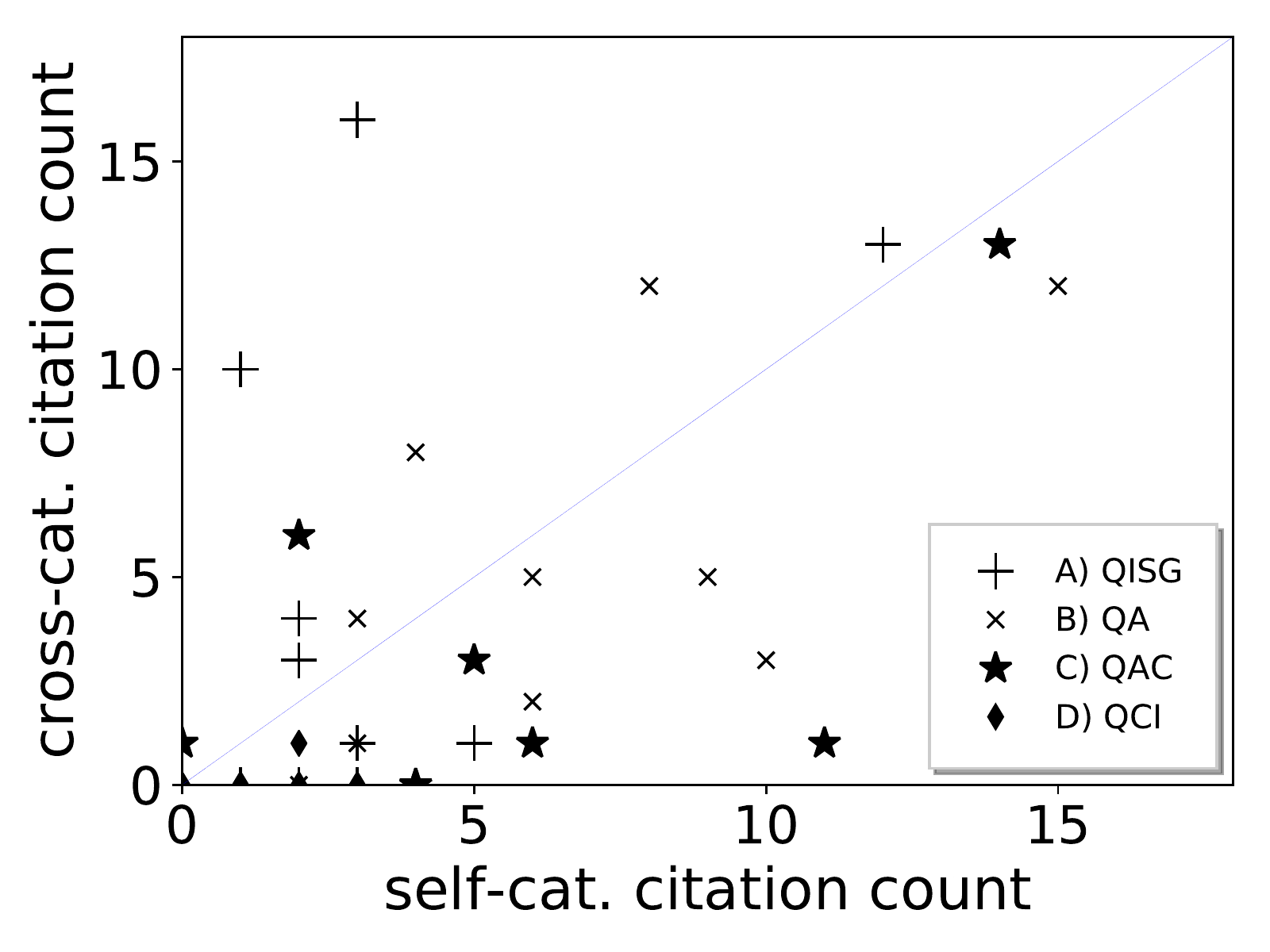}
  \caption{ Correlation plot of cross-category vs self-category citations for each of the selected publications received from among the rest, given in Tables~\ref{tab-tableAB} and \ref{tab-tableCD}. The positions of the data points with respect to the $45^{\circ}$ diagonal line show that there are more in/self-category citations than citations from other/cross-categories, statistically justifying the broad categorization of the nodes (publications) into four groups as given in Tables~\ref{tab-tableAB} and \ref{tab-tableCD}.
 }
  \label{fig-inoutcorr}
 \end{figure}

 \begin{figure}[!htbp]
 \includegraphics[width=9cm,height=6cm]{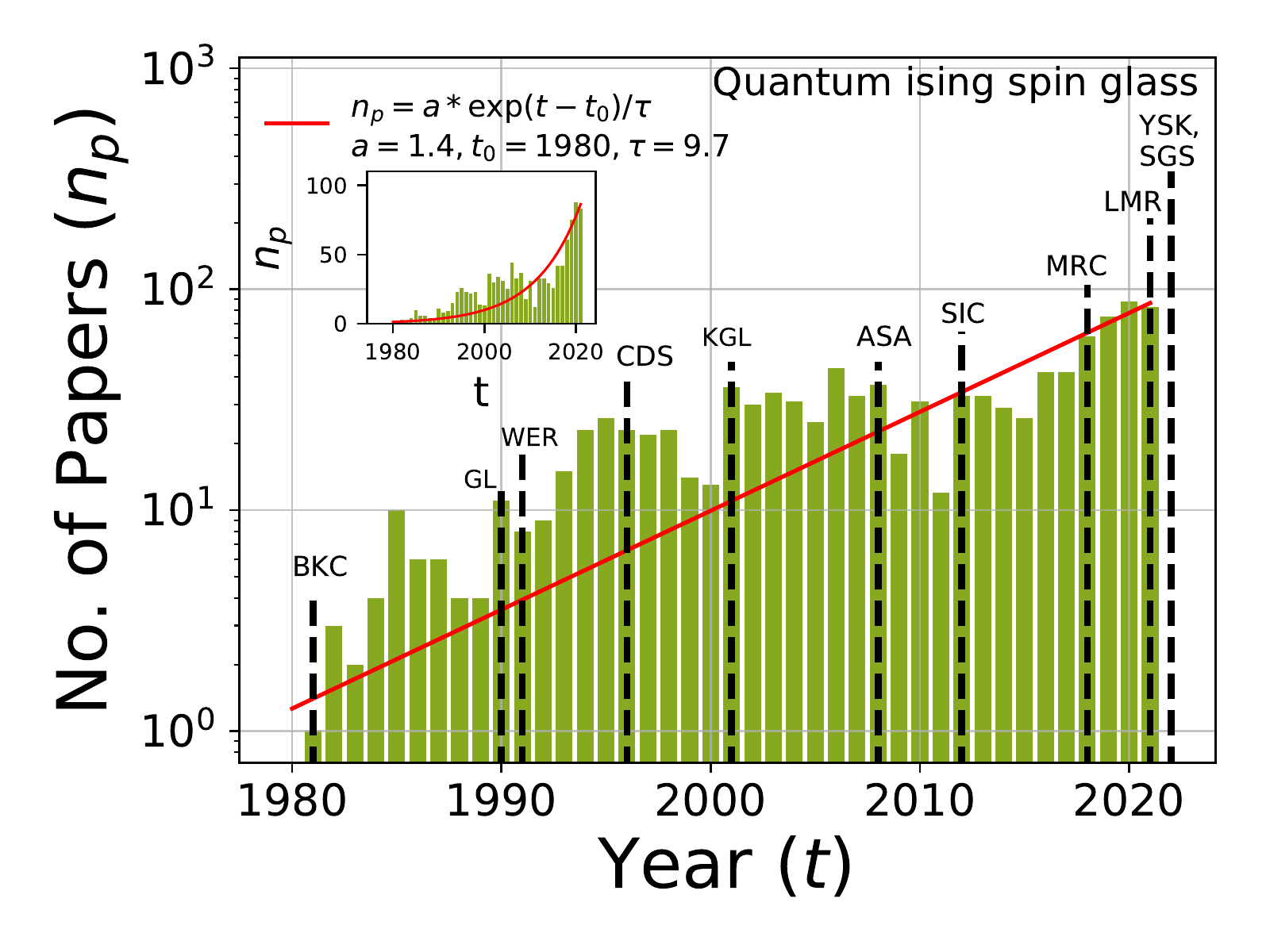}
  \caption{ Annual number of publications ($n_p$) searched from Google scholar (access-date: 8-April-2022) with the exact phrase ``quantum spin glass''  and the additional words ``transverse field'' in the document title or abstract starting from 1980 ($t=t_0$) up to 2021. An exponential growth in $n_p$ is observed over these years. The data fits to $n_p\sim \exp{[(t-t_0)/\tau]}$ with $\tau\simeq9.7$ years. We have marked some of the notable events: BKC denotes Chakrabarti~1981~\cite{101}; GL denotes Goldschmidt, Lai~1990~\cite{38}; WER denotes Wu and Ellman, Rosenbaum et al.~1991~\cite{4}; CDS denotes Chakrabarti, Dutta and  Sen~1996~\cite{102}; KGL denotes Kao, Grest and Levin et al.~2001~\cite{103}; ASA denotes Ancona-Torres, Silevitch, Aeppli  et al.~2008~\cite{48}; SIC denotes Suzuki, Inoue and  Chakrabarti~2012~\cite{104}; MRC denotes Mukherjee, Rajak and  Chakrabarti~2018~\cite{105}; LMR denotes Leschke, Manai and Ruder et al.~2021~\cite{106};   YSK denotes Yaacoby, Schaar, Kellerhals et al.~2022~\cite{107}; SGS denotes Schindler, Guaita, Shi et al.~2022~\cite{108}. The inset shows direct functional relationship between $n_p$ and $t$ for category A from 1980 to 2021. 
 }
  \label{fig-spinglass}
 \end{figure}

\section{Research network and study of its growth}
 \label{2}
 
In this section we study the research network (in subsection~\ref{2}\ref{2A}) containing the nodes (publications), their edges (citations) and also analyze the network growth (in subsection~\ref{2}\ref{2B}). The dataset is obtained using Google scholar (public, popular search engine for scholarly literature across a wide variety of disciplines and sources) and access period of the dataset is 8 to 10 April, 2022. The year of publications of these research documents (nodes/publications) in the dataset range between 1980 and 2021.

\subsection{Research Network}
\label{2A}

In this subsection, we study the searched publications on ``Quantum Annealing Computation and Information'' as a research network that consists of hundred top cited nodes (publications) subdivided into four categories (twenty five in each) as: A) Quantum/Transverse Field Spin Glass Model, B) Quantum Annealing, C) Quantum Adiabatic Computation and D) Quantum Computation Information; see Table~\ref{tab-tableAB}~and~\ref{tab-tableCD}. The network is directed i.e. within network, a node (publication) if receives an edge (citation) from some node doesn't cite or link that node back; the network is shown in Fig.~\ref{fig-timeline}. The order of the network is hundred and the size of the network is nearly about three hundred and fifty considering the edges as citations from latest (by year of publication) ten nodes in each of the categories to any nodes within the network. In Table~\ref{tab-tableinout}, we list top ten publications from the studied research network, that either received maximum citations (indegree), or cited (outdegree) maximum number of nodes of the network among other references; e.g. among the hundred top cited publications, node C1 is cited by 22 other nodes (publications) of the studied network and on the other hand, node C4 has cited 29 nodes from the research network among others. In Fig.~\ref{fig-inoutcorr}, we also show the plot of correlations between cross-category vs self-category citation strength (i.e., how much citations any chosen node has given to other-category nodes vs the self-category nodes) using the references of the latest ten publications from each of the four research categories as given in Tables~\ref{tab-tableAB} and \ref{tab-tableCD}. This network gives an idea about how research among the various component of ``Quantum Computing'' is going on at present time.

   
\begin{figure}[!htbp]
 \includegraphics[width=9cm,height=6cm]{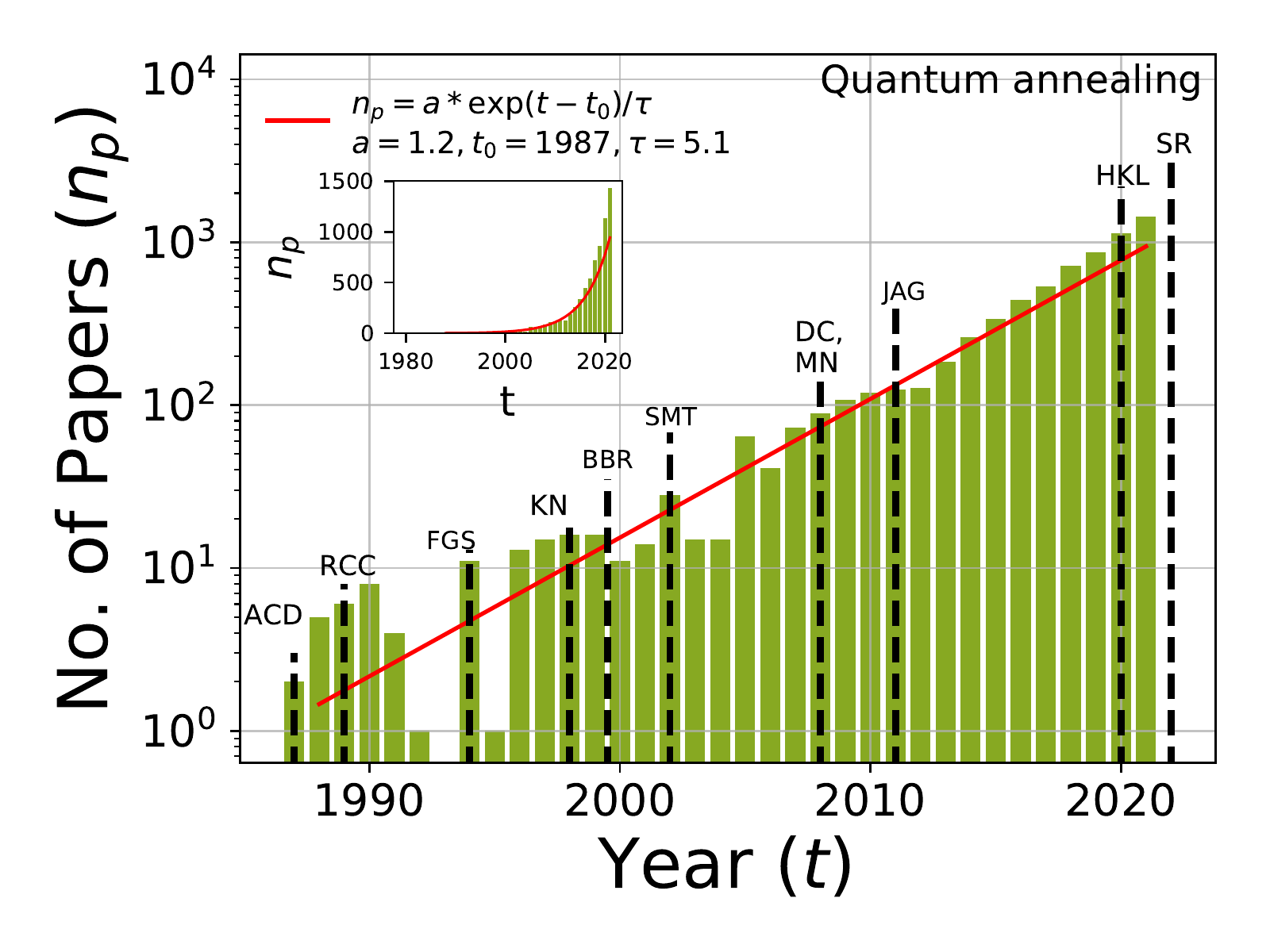}
  \caption{ Annual number of publications ($n_p$) searched from Google scholar (access-date: 8-April-2022) with the exact phrase ``quantum annealing'' in the document title or abstract starting from 1987 ($t=t_0$) upto 2021. An exponential growth in $n_p$ is observed over these years. The data is fitted to $n_p\sim \exp{[(t-t_0)/\tau]}$ with $\tau\simeq5.1$ years. We have marked some of the notable events: ACD denotes Apolloni, Cesa-Bianchi and De Falco~1990~\cite{109}; RCC denotes Ray, Chakrabarti and Chakrabarti~1989~\cite{3}; FGS denotes Finnila, Gomez, Sebenik et al.~1994~\cite{11}; KN denotes Kadowaki and Nishimori~1998~\cite{5}; BBR denotes Brooke, Bitko, Rosenbaum et al.~1999~\cite{15}; SMT denotes Santoro, Marto{\v{n}}{\'a}k, Tosatti et al.~2002~\cite{7}; DC denotes Das and Chakrabarti~2008~\cite{13}; MN denotes Morita and Nishimori~2008~\cite{31}; JAG denotes Johnson, Amin, Gildert et al.~2011~\cite{6}; HKL denotes Hauke, Katzgraber, Lechner et al.~2020~\cite{110}; SR denotes Starchl and Ritsch~2022~\cite{111}. The inset shows direct functional relationship between $n_p$ and $t$ for category B from 1987 to 2021.}
  \label{fig-annealing}
 \end{figure}

  \begin{figure}[!htbp]
 \includegraphics[width=9cm,height=6cm]{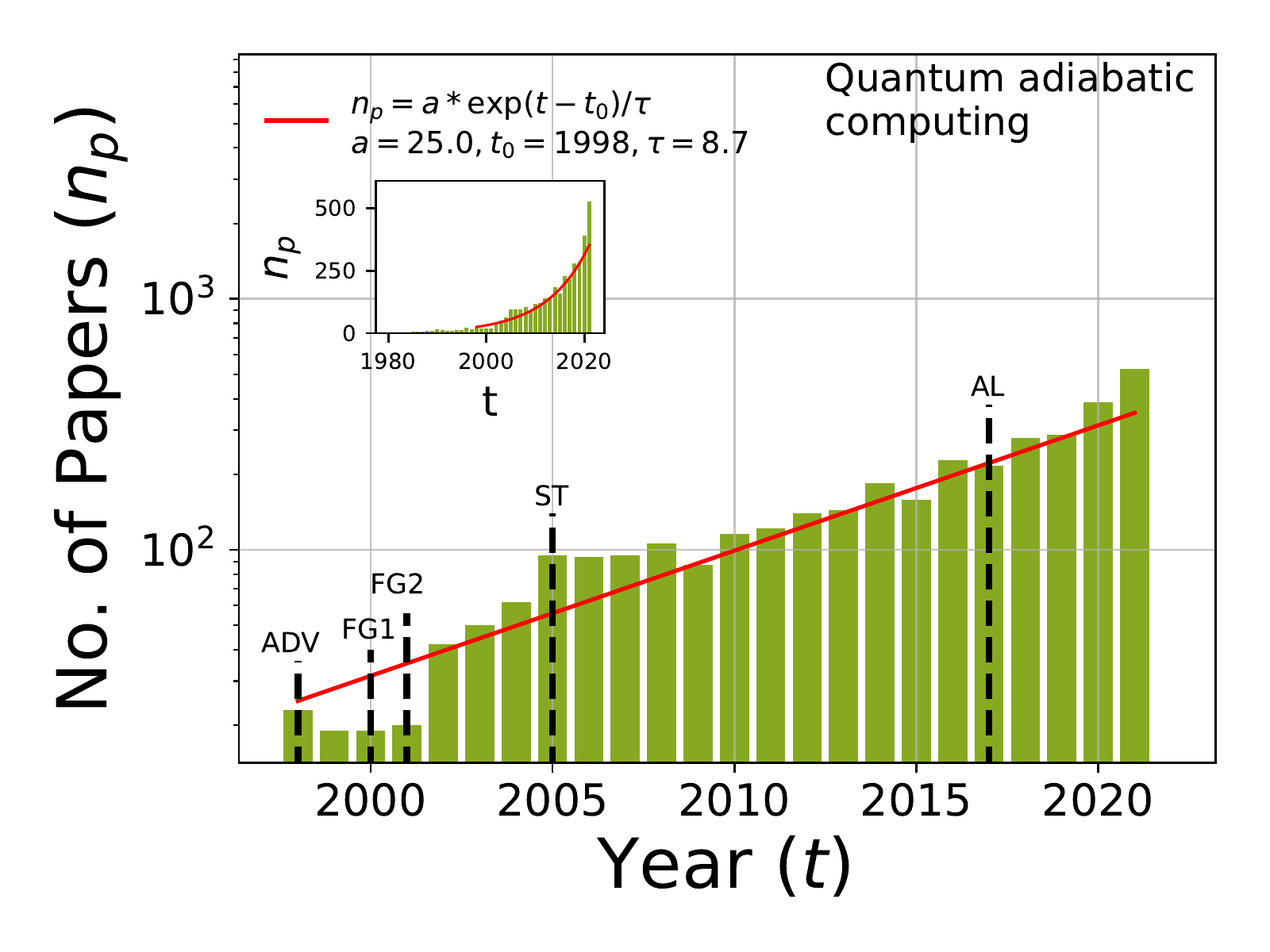}
  \caption{ Annual number of publications ($n_p$) searched from Google scholar (access-date: 8-April-2022) with the exact phrase ``quantum-adiabatic comput'' [where ``comput'' stands for computer, computing or computation] in the document title or abstract starting from 1998 ($t=t_0$) upto 2021. An exponential growth in $n_p$ is observed over these years. The data is fitted to $n_p\sim \exp{[(t-t_0)/\tau]}$ with $\tau\simeq8.7$ years. We have marked some of the notable events: ADV denotes Averin~1998~\cite{64}; FG1 denotes Farhi, Goldstone, Gutmann et al. ~2000~\cite{54} in arxiv; FG2 denotes Farhi, Goldstone, Gutmann et al.~2001~\cite{52}; ST denotes Santoro and Tosatti~2006~\cite{17}; AL denotes Albash and Lidar~2018~\cite{58}; The inset shows direct functional relationship between $n_p$ and $t$ for category C since 1998 upto 2021.}
  \label{fig-adiabatic}
 \end{figure}

\subsection{ Growth Behaviour}
\label{2B}
Here, we study the growth statistics of the annual number of publications $n_p$ for all the four categories (A to D). The growth of $n_p$ with time $t$ (year) is fitted to 
\begin{equation}\label{eq1}
n_p=a*\exp{[(t-t_0)/\tau]},
\end{equation}
\noindent
and scaling time $\tau$ is estimated for each of the categories.\\

 \begin{figure}[!htbp]
 \includegraphics[width=9cm,height=6cm]{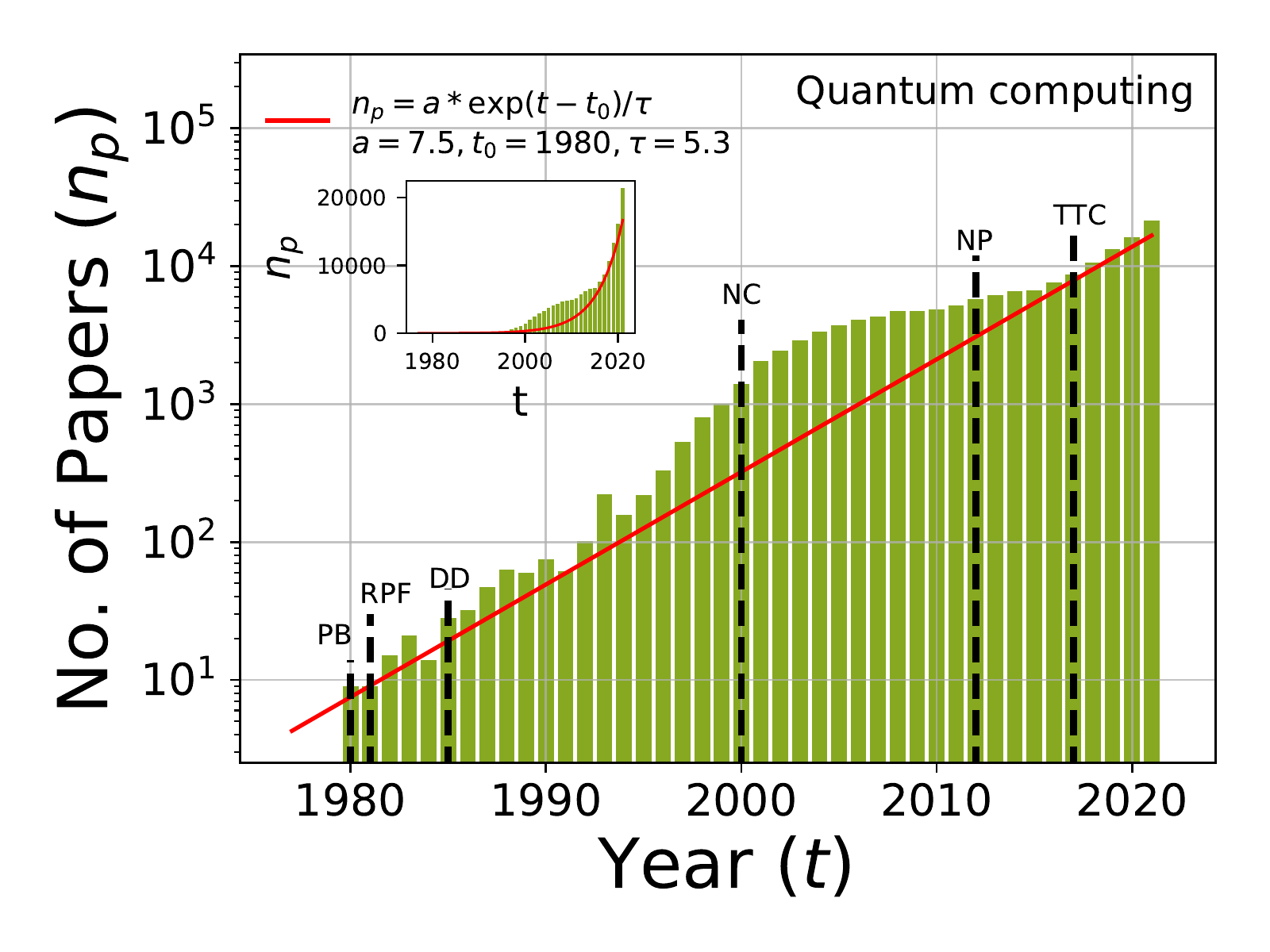}
  \caption{ Annual number of publications ($n_p$) searched from Google scholar (access-date: 8-April-2022) with the exact phrase ``quantum-computing'' in the document title or abstract starting from 1980 ($t=t_0$) upto 2021. An exponential growth in $n_p$ is observed over these years. The data is fitted to $n_p\sim \exp{[(t-t_0)/\tau]}$ with $\tau\simeq5.3$ years. We have marked some of the events: 
  PB denotes Benioff~1980~\cite{112}; RPF denotes Feynman~1982~\cite{2}; DD denotes Deutsch~1985~\cite{113}; NC denotes Nielsen and Chuang~2002~\cite{53}; NP denotes the Nobel Prize (physics) in October, 2012 that Haroche and Winerland received together for their work which ``hold promise for the creation of practical quantum computers''~\cite{114}; TTC denotes Tanaka, Tamura and Chakrabarti~2017~\cite{115}; The inset shows direct functional relationship between $n_p$ and $t$ for category D from 1980 to 2021.}
  \label{fig-computing}
 \end{figure}

   \textbf{A) Quantum/Transverse Field Spin Glass Model (QISG)}
  
  We searched for annual $n_p$ with the exact phrase ``quantum spin glass''  and the additional words ``transverse field'' in the document title or abstract starting from 1980 to 2021. An exponential relationship is observed between $n_p$ and $t$ which fits as Eq.~\ref{eq1} where $t_0=1980$ and $\tau\simeq9.7$, see Fig.~\ref{fig-spinglass}. \\

 \textbf{B) Quantum Annealing (QA)}

We  searched for annual $n_p$ with the exact phrase ``quantum annealing'' in the document title or abstract starting from 1987 to 2021. An exponential relationship is observed between $n_p$ and $t$ which fits as Eq.~\ref{eq1} where $t_0=1987$ and $\tau\simeq5.1$, see Fig.~\ref{fig-annealing}.\\

  \textbf{C) Quantum Adiabatic Computation (QAC)}

 We searched for annual $n_p$ with the exact phrase ``quantum-adiabatic comput'' [where ``comput'' stands for computer, computing or computation] in the document title or abstract starting from 1998 to 2021. An exponential relationship is observed between $n_p$ and $t$ which fits as Eq.~\ref{eq1}  where $t_0=1998$ and  $\tau\simeq8.7$, see Fig.~\ref{fig-adiabatic}.\\


\textbf{D) Quantum Computation Information (QCI)}

We searched for annual $n_p$ with the exact phrase ``quantum-computing'' in the document title or abstract starting from 1980 to 2021. An exponential relationship is observed between $n_p$ and $t$ which fits as Eq.~\ref{eq1} where $t_0=1980$ and $\tau\simeq5.3$, see Fig.~\ref{fig-computing}.\\

 \begin{table}[!htbp]
    \caption{Fitting parameters for Eq.(~\ref{eq1})}
    \label{tab-tablefitparam}
   \begin{tabular}{|l|c|c|c|}
    \hline
        category (research topic) & $a$  & $t_0$ (calendar year) & $\tau$ (years) \\
		\hline     
         A (QISG) & 1.4 & 1980 & 9.7 \\
        \hline
         B (QA) & 1.2 & 1987 & 5.1 \\
        \hline
         C (QAC) & 25.0 & 1998 & 8.7 \\
        \hline
         D (QCI) & 7.5 & 1980 & 5.3 \\
        \hline
        all categories together (A,B,C and D) & 8.5 & 1980 & 5.3 \\
        \hline
		\end{tabular}
\end{table}

 \section{Summary \& Discussion}

Here, we study (in section~\ref{2}\ref{2A}) the structure
of the research  network containing hundred
top cited nodes (publications) from 1980 up to
2021 on the broad topic ``Quantum Annealing
Computation and Information'', searched from
Google Scholar. The searched publications (papers, books) are
categorized into four sub-categories : A)
Quantum/Transverse Field Spin Glass Model,
B) Quantum Annealing,  C) Quantum Adiabatic
Computation and D) Quantum Computation
Information; each consisting  of twenty-five
publications. The citation links between the nodes
of the entire network (of hundred nodes) are
also found. The network is shown in Fig.~\ref{fig-timeline},
and the Google Scholar citations of the hundred
nodes of the network are given in Tables~\ref{tab-tableAB}~and~\ref{tab-tableCD}.
The size of the network (containing the nodes and the
links) is nearly three hundred and fifty.
We considered (see~Table~\ref{tab-tableinout}) the top ten nodes in each category having maximum links (in or out) within the network. 
Apart from self-connections within the
sets (categories), major connections are
seen to flow as AB, AC, BC, BA, CB, CD,
CA and DC. Indeed, the link statistics,
within each categories, and those
across the categories are given in Tables~\ref{tab-tableAB},~\ref{tab-tableCD} and \ref{tab-tableinout}. From Fig.~\ref{fig-inoutcorr},  one can clearly see that the citation correlations within each category is more than cross-category correlations justifying the categorization of the topics as given in Tables~\ref{tab-tableAB} and \ref{tab-tableCD}. 


As may be seen in Figs.~\ref{fig-spinglass},~\ref{fig-annealing},~\ref{fig-adiabatic} and \ref{fig-computing}, for each
category, we have marked a few notable
publications within these forty plus years
(1980-1022). For category A; in Fig.~\ref{fig-spinglass}, we noted the
publications by Chakrabarti 1981~\cite{101} for
the first study of the quantum phase transition 
behavior  of the Mattis-type and the
Edwards-Anderson Ising Spin glasses which,
along  with their further extensions~\cite{3},
were questioned in the theoretical study~\cite{38}
 by Goldschmidt and Lai in 1990, while
supporting experimental evidences were
reported in the pioneering study~\cite{4} in
1991 by Wu, Ellman, Rosenbaum, Aeppli and
Reich. The first monograph on the quantum
phase transitions in transverse Ising spin
(including glass) models was published by
Chakrabarti, Dutta and Sen in 1996~\cite{102}.
The experimental study was further
extended and analyzed by Kao, Grest,
Levin, Brooke, Rosenbaum and Aeppli in
2001~\cite{103}, and further in 2008~\cite{48} by
Ancona-Torres, Silevitch, Aeppli and
Rosenbaum. Next extended monograph
on quantum phase transition in Ising
models and glasses~\cite{104} by Suzuki, Inoue
and Chakrabarti was published in 2012.
Mukherjee, Rajak and Chakrabarti~\cite{105}
published further numerical results
supporting the existence of ergodic
region of the spin glass phase in
transverse Ising Sherrington-Kirkpatrick
model in 2018, while the analytical
study~\cite{106} in 2021 by Leschke, Manai and
Ruder found some support for such tunneling
induced ergodicity in the model. Two recent
papers in 2022 on this problem, namely that~\cite{107}
 by Yaacoby, Schaar, Kellerhals, Raz,
Hermelin and Pugatch and that~\cite{108} by
Schindler, Guaita, Shi, Demler and
Cirac throw some encouraging new lights
on this point. For category B, in Fig.~\ref{fig-annealing},
similarly marked publications starting with the
1988-1990 one by Apolloni, Cesa-Bianchi and
De Falco~\cite{109} having put the term ``quantum
annealing'' in the title of the conference
publication. The seminal 1989 paper by Ray,
Chakrabarti and Chakrabarti~\cite{3} indicated
that quantum fluctuations can help explore
rugged energy landscapes of the classical
Ising spin glasses by escaping from local
minima (having tall but thin barriers) using
tunneling. Finnila, Gomez, Sebenik, Stenson
and Doll~\cite{11} proposed it for search of the
ground state in large molecules in 1994.
Kadowaki and Nishimori~\cite{5} in 1998 first
formulated  and numerically demonstrated
its advantages in Ising glass-like systems.
Brooke, Bitko, Rosenbaum and Aeppli~\cite{15} 
demonstrated experimentally the advantages
in mixed magnets in 1999.  Santoro, Marton
and Tosatti~\cite{7} demonstrated the same in 2002.
First major review on quantum annealing was
published in 2008 by Das and Chakrabarti~\cite{13}.
Its mathematical structure was reviewed by
Morita and Nishimori in 2008~\cite{31}. Following
a major development and publication~\cite{6} by
Johnson and coworkers in 2011, the D-wave
quantum annealing computers became available
in the market. Hauke, Katzgraber, Lechner,
Nishimori and Oliver reviewed the latest
developments. Starchl and Ritsch~\cite{111} in 2022
first reviewed briefly the developments
(notably identifying ref.~\cite{3} by Ray et al. as
``the earliest work in laying the foundation of
quantum annealing'') and  demonstrated
theoretically the efficiencies of photonic
annealers. For category C, in Fig.~\ref{fig-adiabatic}, we
highlighted the 1998 publication by Averin~\cite{64}
for proposing first the adiabatic quantum
computation with Cooper pairs. Next, in two
successive pioneering publications in 2000 and
2001 by Farhi, Goldstone, Gutmann and Sipser~\cite{54}, and by Farhi, Goldstone, Gutmann, Lapan,
Lundgren and Preda~\cite{52}, the quantum adiabatic
computation was precisely formulated.  Santoro
and Tosatti reviewed the developments in
2006~\cite{17} and  Albash and Lidar~\cite{58} in 2018.
For category D, in Fig.~\ref{fig-computing}, we have
marked the classic publications~\cite{112},~\cite{2} and~\cite{113}  by  Benioff in  1980, by Feynman in 1982
and by Deutsch in  1985 respectively. The
classic book~\cite{53} on quantum computation and
information by Nielsen and Chuang appeared in
2002. The 2012 Physics Nobel Prize went to
Haroche and Winerland for their work which
``hold promise for the creation of practical
quantum computers''~\cite{114}. A theoretical
compendium~\cite{115} by  Tanaka, Tamura and
Chakrabarti appeared in 2017.

For the study of the growth of the research
literature in the general topic ``Quantum
Annealing Computation and Information'', we
again studied (in section~\ref{2}\ref{2B}), the annual
growth  rate $n_p$ of publications in all
the four categories (A-D) with time $t$ (in
years) within the period 1980 to 2021. All the 
data were collected during 8 to 10
April 2022, using the Google scholar.
This annual Number of Publications ($n_p$) in
each of these categories were the fitted to 
Eq.~\ref{eq1}, namely $n_p = a*exp
[(t - t_0)/{\tau}]$ and estimated the growth
scaling time $\tau$ for each of the categories (see Figs.~\ref{fig-spinglass},~\ref{fig-annealing},~\ref{fig-adiabatic},~\ref{fig-computing}); category-wise fitting parameters are given in Table~\ref{tab-tablefitparam}. The fitting process shows
the initial number of publications ($a = n_p$ at $t
= t_0$ in Eq.~\ref{eq1}) to vary in
the range 1 to 20 and the estimated value
of $t_0$ compares well with the first notable
publication of the topic or category (see the
indicated years of the notable publication, 
 Nobel Prize related to quantum computing), as shown in~Figs.~\ref{fig-spinglass},~\ref{fig-annealing},~\ref{fig-adiabatic},~\ref{fig-computing}.

Overall, the literature in quantum annealing
and computing is growing  exponentially (from
about $10^1$ at $t = t_0 \sim 1980$ to order
$10^2$ - $10^4$ at $t=2021$) with time
and while for categories like quantum spin glass
(A) and quantum adiabatic computation (C), the
typical growth time $\tau$ (during which
the number of publications grows by a factor
$e \simeq 2.7$) is about 10 years, that
for the categories quantum annealing (B) and
quantum computing (D) is seen to be about 5
years (see Figs.~\ref{fig-spinglass},~\ref{fig-annealing},~\ref{fig-adiabatic},~\ref{fig-computing}). This 
clearly indicates remarkable activity in the research fields Quantum Annealing (category B) and Quantum Computation (category D).  It may be mentioned that the growth behaviour of the number of publications for all these four categories together also fits Eq.~\ref{eq1} with $t_0=1980$ and $\tau\simeq5.3$. This growth may be compared with that of a very active contemporary field, namely quantitative economics including econophysics (see~\cite{116}),  
 having growth scaling time $\tau$ of about 8 years.

 \dataccess{All data used in this work are publicly available.}

 
 \funding{Author received no funding for this study.}
 
\ack{Author is grateful to Bikas K. Chakrabarti for suggesting this study and to Atanu Rajak, Sudip Mukherjee for their useful comments and suggestions on the manuscript. The author is thankful to two anonymous referees for their important comments and suggestions. }


\begin{thebibliography}{9}


\bibitem{1} Kirkpatrick, Scott and Gelatt Jr, C Daniel and Vecchi, Mario P.,  
\textit{Optimization by simulated annealing},
Science, 
\textbf{220; 671--680} (1983).


\bibitem{2} Feynman, Richard P,   
\textit{Simulating physics with computers}, International Journal of Theoretical Physics,  
\textbf{21; 467-488} (1982).


\bibitem{3} Ray, Purusattam and Chakrabarti, Bikas K and Chakrabarti, Arunava,    
\textit{Sherrington-Kirkpatrick model in a transverse field: Absence of replica symmetry breaking due to quantum fluctuations}, Physical Review B, \textbf{39; 11828} (1989).


\bibitem{4} Wu, Wenhao and Ellman, B and Rosenbaum, Thomas Felix and Aeppli, Gabriel and Reich, Daniel H,      \textit{From classical to quantum glass}, Physical review letters,  \textbf{67; 2076} (1991).


\bibitem{5} Kadowaki, Tadashi and Nishimori, Hidetoshi,    
\textit{Quantum annealing in the transverse Ising model}, Physical Review E,  \textbf{58; 5355} (1998).



\bibitem{6} Johnson, Mark W and Amin, Mohammad HS and Gildert, Suzanne and Lanting, Trevor and Hamze, Firas and Dickson, Neil and Harris, Richard and Berkley, Andrew J and Johansson, Jan and Bunyk, Paul and others,    
\textit{Quantum annealing with manufactured spins}, Nature,   \textbf{473; 194--198} (2011).


\bibitem{7} Santoro, Giuseppe E and Marton{\'a}k, Roman and Tosatti, Erio and Car, Roberto,      
\textit{Theory of quantum annealing of an Ising spin glass},  Science, \textbf{295; 2427--2430} (2002).
 
 
\bibitem{8}  Dutta, Amit and Aeppli, Gabriel and Chakrabarti, Bikas K and Divakaran, Uma and Rosenbaum, Thomas Felix and Sen, Diptiman,  
\textit{Quantum phase transitions in transverse field spin models: from statistical physics to quantum information}, Cambridge University Press, Cambridge, (2015).


\bibitem{9} Boixo, Sergio and R{\o}nnow, Troels F and Isakov, Sergei V and Wang, Zhihui and Wecker, David and Lidar, Daniel A and Martinis, John M and Troyer, Matthias,    \textit{Evidence for quantum annealing with more than one hundred qubits}, Nature physics,  \textbf{10;218--224} (2014).


\bibitem{10}  Cragg, Dinah M and Sherrington, David and Gabay, Marc,     \textit{Instabilities of an m-Vector Spin-Glass in a Field}, Physical Review Letters,  \textbf{49;158} (1982).
 
 
\bibitem{11}  Finnila, Aleta Berk and Gomez, MA and Sebenik, C and Stenson, Catherine and Doll, Jimmie D,   \textit{Quantum annealing: A new method for minimizing multidimensional functions},  Chemical physics letters,  \textbf{219; 343--348} (1994).


\bibitem{12} Strack, Philipp and Sachdev, Subir,    \textit{Dicke quantum spin glass of atoms and photons}, Physical review letters,  \textbf{107; 277202} (2011).


\bibitem{13} Das, Arnab and Chakrabarti, Bikas K,       \textit{Colloquium: Quantum annealing and analog quantum computation}, Reviews of Modern Physics,  \textbf{80; 1061} (2008).
 
 
\bibitem{14} Heim, Bettina and R{\o}nnow, Troels F and Isakov, Sergei V and Troyer, Matthias,    \textit{Quantum versus classical annealing of Ising spin glasses},  Science,  \textbf{348; 215--217} (2015).



\bibitem{15} Brooke, J and Bitko, David and Rosenbaum Thomas Felix and Aeppli, Gabriel,    \textit{Quantum annealing of a disordered magnet}, Science,  \textbf{284; 779--781} (1999).


\bibitem{16}  Rieger, Heiko and Young, A Peter,    \textit{Zero-temperature quantum phase transition of a two-dimensional Ising spin glass},  Physical review letters, \textbf{72; 4141} (1994).

 
\bibitem{17}  Santoro, Giuseppe E and Tosatti, Erio,  \textit{Optimization using quantum mechanics: quantum annealing through adiabatic evolution}, Journal of Physics A: Mathematical and General,   \textbf{39; R393} (2006).


\bibitem{18} Ye, Jinwu and Sachdev, Subir and Read, Nicholas,    \textit{Solvable spin glass of quantum rotors}, Physical review letters,  \textbf{70; 4011} (1993).


\bibitem{19} Lanting, Trevor and Przybysz, Anthony J and Smirnov, A Yu and Spedalieri, Federico M and Amin, Mohammad H and Berkley, Andrew J and Harris, Richard and Altomare, Fabio and Boixo, Sergio and Bunyk, Paul and others      \textit{Entanglement in a quantum annealing processor}, Physical Review X,   \textbf{4; 021041} (2014).


\bibitem{20} Read, Nicholas and Sachdev, Subir and Ye, Jinwu,   \textit{Landau theory of quantum spin glasses of rotors and Ising spins}, Physical Review B,   \textbf{52; 384} (1995).

\bibitem{21} Bunyk, Paul I and Hoskinson, Emile M and Johnson, Mark W and Tolkacheva, Elena and Altomare, Fabio and Berkley, Andrew J and Harris, Richard and Hilton, Jeremy P and Lanting, Trevor and Przybysz, Anthony J and others,    \textit{Architectural considerations in the design of a superconducting quantum annealing processor}, IEEE Transactions on Applied Superconductivity, \textbf{24; 1--10} (2014).

\bibitem{22} Harris, Richard and Sato, Y and Berkley, Andrew J and Reis, M and Altomare, Fabio and Amin, MH and Boothby, Kelly and Bunyk, P and Deng, C and Enderud, Colin and Huang, Shuiyuan and Hoskinson, E and Johnson, MW and Ladizinsky, E and Ladizinsky, N and Lanting, T and Li, R and Medina, T and Molavi, R and Neufeld, R and Oh, T and Pavlov, I and Perminov, I and Poulin-Lamarre, Gabriel and Rich, Christopher and Smirnov, Anatoly and Swenson, L and Tsai, Nicholas and Volkmann, M and Whittaker, Jed and Yao, J,      \textit{Phase transitions in a programmable quantum spin glass simulator}, Science,  \textbf{361; 162--165} (2018).

\bibitem{23} Perdomo-Ortiz, Alejandro and Dickson, Neil and Drew-Brook, Marshall and Rose, Geordie and Aspuru-Guzik, Al{\'a}n,   \textit{Finding low-energy conformations of lattice protein models by quantum annealing}, Scientific reports,   \textbf{2; 1--7} (2012).

\bibitem{24}  Guo, Muyu and Bhatt, Ravindra N and Huse, David A,   \textit{Quantum critical behavior of a three-dimensional Ising spin glass in a transverse magnetic field}, Physical review letters, \textbf{72; 4137} (1994).

\bibitem{25} Boixo, Sergio and Albash, Tameem and Spedalieri, Federico M and Chancellor, Nicholas and Lidar, Daniel A,      \textit{Experimental signature of programmable quantum annealing}, Nature communications,  \textbf{4; 1--8} (2013).

\bibitem{26}  Miller, Jonathan and Huse, David A,   \textit{Zero-temperature critical behavior of the infinite-range quantum Ising spin glass},  Physical review letters,  \textbf{70; 3147} (1993).


\bibitem{27} Das, Arnab and Chakrabarti, Bikas K    \textit{Quantum annealing and related optimization methods}, Springer Science \& Business Media, \textbf{679} (2005).


\bibitem{28}  Thill, MJ and Huse, David A,     \textit{Equilibrium behaviour of quantum Ising spin glass}, Physica A: Statistical Mechanics and its Applications,   \textbf{214; 321--355} (1995).


\bibitem{29}  Dickson, Neil G and Johnson, Mark W and Amin, Mohammad and Harris, R and Altomare, F and Berkley, AJ and Bunyk, Paul and Cai, J and Chapple, EM and Chavez, P and Cioata, F and Cirip, T and deBuen, P and Drew-Brook, M and Enderud, C and Gildert, S and Hamze, F and Hilton, JP and Hoskinson, E and Karimi, K and Ladizinsky, E and  Ladizinsky, N and Lanting, T and Mahon, T and Neufeld, R and Oh, T and Perminov, I and Petroff, C and Przybysz, A and Rich, C and Spear, P and Tcaciuc, A and Thom, MC and Tolkacheva, E and Uchaikin, S and Wang, J and Wilson, AB and Merali, Z and Rose, G ,  \textit{Thermally assisted quantum annealing of a 16-qubit problem}, Nature communications,   \textbf{4; 1--6} (2013).


\bibitem{30}  Georges, Antoine and Parcollet, Olivier and Sachdev, Subir,   \textit{Mean field theory of a quantum Heisenberg spin glass}, Physical review letters,  \textbf{85; 840} (2000).


\bibitem{31}  Morita, Satoshi and Nishimori, Hidetoshi,     \textit{Mathematical foundation of quantum annealing}, Journal of Mathematical Physics,  \textbf{49; 125210} (2008).


\bibitem{32}  Rieger, Heiko and Young, A. Peter,  \textit{Griffiths singularities in the disordered phase of a quantum Ising spin glass},  Physical Review B,  \textbf{54; 3328} (1996).


\bibitem{33}  Adachi, Steven H and Henderson, Maxwell P,   \textit{Application of quantum annealing to training of deep neural networks},  arXiv preprint arXiv:1510.06356 (2015).


\bibitem{34}  Hen, Itay and Job, Joshua and Albash, Tameem and R{\o}nnow, Troels F and Troyer, Matthias and Lidar, Daniel A,     \textit{Probing for quantum speedup in spin-glass problems with planted solutions}, Physical Review A,   \textbf{92; 042325} (2015).


\bibitem{35} Pudenz, Kristen L and Albash, Tameem and Lidar, Daniel A,   \textit{Error-corrected quantum annealing with hundreds of qubits}, Nature communications,    \textbf{5; 1--10} (2014).


\bibitem{36}  Laumann, Christopher R and Pal, A and Scardicchio, A,   \textit{Many-body mobility edge in a mean-field quantum spin glass}, Physical review letters,  \textbf{113; 200405} (2014).


\bibitem{37}  Marto{\v{n}}{\'a}k, Roman and Santoro, Giuseppe E and Tosatti, Erio,     \textit{Quantum annealing of the traveling-salesman problem}, Physical Review E, \textbf{70; 057701} (2004).


\bibitem{38} Goldschmidt, Yadin Y and Lai, Pik-Yin,   \textit{Ising spin glass in a transverse field: Replica-symmetry-breaking solution}, Physical review letters,    \textbf{64; 2467} (1990).


\bibitem{39}  Lechner, Wolfgang and Hauke, Philipp and Zoller, Peter,   \textit{A quantum annealing architecture with all-to-all connectivity from local interactions}, Science advances,  \textbf{1; e1500838} (2015).


\bibitem{40}  Guo, Muyu and Bhatt, RN and Huse, David A,     \textit{Quantum Griffiths singularities in the transverse-field Ising spin glass},  Physical Review B, \textbf{54; 3336} (1996).

\bibitem{41}  Glassy chimeras could be blind to quantum speedup: Designing better benchmarks for quantum annealing machines,  \textit{Glassy chimeras could be blind to quantum speedup: Designing better benchmarks for quantum annealing machines}, Physical Review X,   \textbf{4; 021008} (2014).


\bibitem{42}   Goldschmidt, Yadin Y,  \textit{Solvable model of the quantum spin glass in a transverse field},  Physical Review B, \textbf{41; 4858} (1990).


\bibitem{43}  McGeoch, Catherine C,     \textit{Adiabatic quantum computation and quantum annealing: Theory and practice}, Synthesis Lectures on Quantum Computing,  \textbf{5; 1--93} (2014).

\bibitem{44} Ishii, Hiroumi and Yamamoto, Tetsuya,   \textit{Effect of a transverse field on the spin glass freezing in the Sherrington-Kirkpatrick model}, Journal of Physics C: Solid State Physics,   \textbf{18; 6225} (1985).


\bibitem{45}  Marto{\v{n}}{\'a}k, Roman and Santoro, Giuseppe E and Tosatti, Erio,   \textit{Quantum annealing by the path-integral Monte Carlo method: The two-dimensional random Ising model}, Physical Review B,  \textbf{66; 094203} (2002).


\bibitem{46}  Yamamoto, Tetsuya and Ishii, Hiroumi,      \textit{A perturbation expansion for the Sherrington-Kirkpatrick model with a transverse field}, Journal of Physics C: Solid State Physics,  \textbf{20; 6053} (1987).

\bibitem{47} Mott, Alex and Job, Joshua and Vlimant, Jean-Roch and Lidar, Daniel and Spiropulu, Maria,   \textit{Solving a Higgs optimization problem with quantum annealing for machine learning}, Nature,   \textbf{550; 375--379} (2017).


\bibitem{48} Ancona-Torres, C and Silevitch, Daniel M and Aeppli, Gabriel and Rosenbaum, Thomas Felix,    \textit{Quantum and classical glass transitions in LiHo x Y 1- x F 4},  Physical review letters, \textbf{101; 057201} (2008).


\bibitem{49} Battaglia, Demian A and Santoro, Giuseppe E and Tosatti, Erio,       \textit{Optimization by quantum annealing: Lessons from hard satisfiability problems}, Physical Review E,  \textbf{71; 066707} (2005).

\bibitem{50} Thirumalai, Devarajan and Li, Qiang and Kirkpatrick, Theodore Ross,   \textit{Infinite-range Ising spin glass in a transverse field}, Journal of Physics A: Mathematical and General,  \textbf{22; 3339} (1989).


\bibitem{51}  Somma, Rolando D and Nagaj, Daniel and Kieferov{\'a}, M{\'a}ria,   \textit{Quantum speedup by quantum annealing}, Physical review letters  \textbf{109; 050501} (2012).


\bibitem{52}   Farhi, Edward and Goldstone, Jeffrey and Gutmann, Sam and Lapan, Joshua and Lundgren, Andrew and Preda, Daniel,    \textit{A quantum adiabatic evolution algorithm applied to random instances of an NP-complete problem}, Science \textbf{292; 472--475} (2001).


\bibitem{53}  Nielsen, Michael A and Chuang, Isaac,  \textit{Quantum computation and quantum information}, American Association of Physics Teachers, (2000).


\bibitem{54}  Farhi, Edward and Goldstone, Jeffrey and Gutmann, Sam and Sipser, Michael,   \textit{Quantum computation by adiabatic evolution}, arXiv preprint quant-ph\/0001106, (2000).


\bibitem{55}  Knill, Emanuel and Laflamme, Raymond and Milburn, Gerald J,     \textit{A scheme for efficient quantum computation with linear optics}, Nature  \textbf{409; 46--52} (2001).

\bibitem{56}  Aharonov, Dorit and Van Dam, Wim and Kempe, Julia and Landau, Zeph and Lloyd, Seth and Regev, Oded,  \textit{Adiabatic quantum computation is equivalent to standard quantum computation},  SIAM review,  \textbf{50; 755-787} (2008).


\bibitem{57} Deutsch, David and Jozsa, Richard,    \textit{Rapid solution of problems by quantum computation}, Proceedings of the Royal Society of London. Series A: Mathematical and Physical Sciences,   \textbf{439; 553--558} (1992).


\bibitem{58}  Albash, Tameem and Lidar, Daniel A,     \textit{Adiabatic quantum computation}, Reviews of Modern Physics,  \textbf{90; 015002} (2018).

\bibitem{59} DiVincenzo, David P,   \textit{The physical implementation of quantum computation},  Fortschritte der Physik: Progress of Physics,  \textbf{48; 771--783} (2000).


\bibitem{60} Childs, Andrew M and Farhi, Edward and Preskill, John,    \textit{Robustness of adiabatic quantum computation}, Physical Review A,  \textbf{65; 012322} (2001).


\bibitem{61} Bennett, Charles H and DiVincenzo, David P,      \textit{Quantum information and computation}, Nature,  \textbf{404; 247--255} (2000).

\bibitem{62} Barends, Rami and Shabani, Alireza and Lamata, Lucas and Kelly, Julian and Mezzacapo, Antonio and Las Heras, Urtzi and Babbush, Ryan and Fowler, Austin G and Campbell, Brooks and Chen, Yu and others,   \textit{Digitized adiabatic quantum computing with a superconducting circuit}, Nature,  \textbf{534; 222--226} (2016).


\bibitem{63} Bouwmeester, Dirk and Zeilinger, Anton,    \textit{The physics of quantum information: basic concepts}, The physics of quantum information,  Springer, \textbf{1--14} (2000).


\bibitem{64} Averin, Dmitri V,  \textit{Adiabatic quantum computation with Cooper pairs}, Solid State Communications,  \textbf{105; 659--664} (1998).

\bibitem{65} Ekert, Artur and Jozsa, Richard,   \textit{Quantum computation and Shor's factoring algorithm}, Reviews of Modern Physics,  \textbf{68; 733} (1996).


\bibitem{66} Jansen, Sabine and Ruskai, Mary-Beth and Seiler, Ruedi,    \textit{Bounds for the adiabatic approximation with applications to quantum computation}, Journal of Mathematical Physics,  \textbf{48; 102111} (2007).


\bibitem{67}  Steane, Andrew,     \textit{Quantum computing}, Reports on Progress in Physics,  \textbf{61; 117} (1998).

\bibitem{68} Choi, Vicky,   \textit{Minor-embedding in adiabatic quantum computation: I. The parameter setting problem}, Quantum Information Processing   \textbf{7; 193--209} (2008).

\bibitem{69} Raussendorf, Robert and Browne, Daniel E and Briegel, Hans J,    \textit{Measurement-based quantum computation on cluster states}, Physical review A,   \textbf{68; 022312} (2003).

\bibitem{70}  Sj{\"o}qvist, Erik and Tong, Dian-Min and Andersson, L Mauritz and Hessmo, Bj{\"o}rn and Johansson, Markus and Singh, Kuldip,     \textit{Non-adiabatic holonomic quantum computation}, New Journal of Physics,  \textbf{14; 103035} (2012).

\bibitem{71} Kitaev, Alexei Yu and Shen, Alexander and Vyalyi, Mikhail N and Vyalyi, Mikhail N,   \textit{Classical and quantum computation}, American Mathematical Soc., \textbf{47} (2002).


\bibitem{72} Van Dam, Wim and Mosca, Michele and Vazirani, Umesh,    \textit{How powerful is adiabatic quantum computation?}, Proceedings 42nd IEEE symposium on foundations of computer science, \textbf{279--287} (2001).


\bibitem{73}  DiVincenzo, David P and Bacon, Dave and Kempe, Julia and Burkard, Guido and Whaley, K Birgitta,     \textit{Universal quantum computation with the exchange interaction}, Nature,  \textbf{408; 339--342} (2000).

\bibitem{74} Mizel, Ari and Lidar, Daniel A and Mitchell, Morgan,   \textit{Simple proof of equivalence between adiabatic quantum computation and the circuit model},  Physical review letters,  \textbf{99; 070502} (2007).


\bibitem{75} Lloyd, Seth and Braunstein, Samuel L.     \textit{Quantum computation over continuous variables},  Springer, \textbf{9--17} (1999).


\bibitem{76} Choi, Vicky,   \textit{Minor-embedding in adiabatic quantum computation: II. Minor-universal graph design}, Quantum Information Processing, Springer,   \textbf{10; 343--353} (2011).

\bibitem{77} Zanardi, Paolo and Rasetti, Mario,   \textit{Holonomic quantum computation}, Physics Letters A,   \textbf{264; 94--99} (1999).


\bibitem{78}  Sarandy, MS and Lidar, DA,   \textit{Adiabatic quantum computation in open systems}, Physical review letters,  \textbf{95; 250503} (2005).


\bibitem{79} Aspuru-Guzik, Al{\'a}n and Dutoi, Anthony D and Love, Peter J and Head-Gordon, Martin,      \textit{Simulated quantum computation of molecular energies}, Science,  \textbf{309; 1704--1707} (2005).

\bibitem{80}  Altshuler, Boris and Krovi, Hari and Roland, J{\'e}r{\'e}mie,  \textit{Anderson localization makes adiabatic quantum optimization fail}, Proceedings of the National Academy of Sciences,   \textbf{107; 12446--12450} (2010).


\bibitem{81} Jones, Jonathan A and Vedral, Vlatko and Ekert, Artur and Castagnoli, Giuseppe,    \textit{Geometric quantum computation using nuclear magnetic resonance}, Nature,  \textbf{403; 869--871} (2000).


\bibitem{82}   Jordan, Stephen P and Farhi, Edward and Shor, Peter W,    \textit{Error-correcting codes for adiabatic quantum computation}, Physical Review A,  \textbf{74; 052322} (2006).

\bibitem{83} Freedman, Michael and Kitaev, Alexei and Larsen, Michael and Wang, Zhenghan,   \textit{Topological quantum computation}, Bulletin of the American Mathematical Society,  \textbf{40; 31--38} (2003).


\bibitem{84} Babbush, Ryan and Love, Peter J and Aspuru-Guzik, Al{\'a}n,    \textit{Adiabatic quantum simulation of quantum chemistry}, Scientific reports,  \textbf{4; 1--11} (2014).


\bibitem{85} Raussendorf, Robert and Harrington, Jim,      \textit{Fault-tolerant quantum computation with high threshold in two dimensions}, Physical review letters,  \textbf{98; 190504} (2007).

\bibitem{86} Rezakhani, Ali T and Kuo, W-J and Hamma, Alioscia and Lidar, Daniel A and Zanardi, Paolo,    \textit{Quantum adiabatic brachistochrone}, Physical review letters,   \textbf{103; 080502} (2009).


\bibitem{87}  Briegel, Hans J and Browne, David E and D{\"u}r, Wolfgang and Raussendorf, Robert and Van den Nest, Maarten,   \textit{Measurement-based quantum computation}, Nature Physics,   \textbf{5; 19--26} (2009).


\bibitem{88}  Lidar, Daniel A and Rezakhani, Ali T and Hamma, Alioscia,     \textit{Adiabatic approximation with exponential accuracy for many-body systems and quantum computation}, Journal of Mathematical Physics,  \textbf{50; 102106} (2009).

\bibitem{89} Duan, Lu-Ming and Guo, Guang-Can,   \textit{Preserving coherence in quantum computation by pairing quantum bits}, Physical Review Letters   \textbf{79; 1953} (1997).

\bibitem{90} Lidar, Daniel A,    \textit{Towards fault tolerant adiabatic quantum computation}, Physical Review Letters,  \textbf{100; 160506} (2008).


\bibitem{91}  Ralph, Timothy C and Gilchrist, Alexei and Milburn, Gerard J and Munro, William J and Glancy, Scott,     \textit{Quantum computation with optical coherent states}, Physical Review A,  \textbf{68; 042319} (2003).

\bibitem{92}  Sarma, Sankar D and Freedman, Michael and Nayak, Chetan,  \textit{Majorana zero modes and topological quantum computation}, NPJ Quantum Information   \textbf{1; 1--13} (2015).


\bibitem{93} Albash, Tameem and Lidar, Daniel A,    \textit{Decoherence in adiabatic quantum computation}, Physical Review A,  \textbf{91; 062320} (2015).


\bibitem{94}   Bennett, Charles H,    \textit{Quantum information and computation}, Physics Today,  \textbf{48; 24--30} (1995).

\bibitem{95} Amin, MHS and Love, Peter J and Truncik, CJS    \textit{Thermally assisted adiabatic quantum computation},   \textbf{100; 060503} (2008).


\bibitem{96} Knill, Emanuel and Laflamme, Raymond and Zurek, Wojciech H,    \textit{Resilient quantum computation}, Science,  \textbf{279; 342--345} (1998).


\bibitem{97} Roland, Jeremie and Cerf, Nicolas J,      \textit{Noise resistance of adiabatic quantum computation using random matrix theory},  \textbf{71; 032330} (2005).

\bibitem{98} Freedman, Michael H and Larsen, Michael and Wang, Zhenghan,   \textit{A modular functor which is universal{\P} for quantum computation}, Communications in Mathematical Physics, Springer,  \textbf{227; 605--622} (2002).


\bibitem{99} Zheng, Shi-Biao,    \textit{Nongeometric conditional phase shift via adiabatic evolution of dark eigenstates: a new approach to quantum computation},  Physical review letters, \textbf{95; 080502} (2005).


\bibitem{100}   Benenti, Giuliano and Casati, Giulio and Strini, Giuliano,    \textit{Principles of Quantum Computation and Information-Volume II: Basic Tools and Special Topics}, World Scientific Publishing Company,  (2007).

\bibitem{101}  Chakrabarti, Bikas K,
\textit{Critical behavior of the Ising spin-glass models in a transverse field}, Physical Review B,   \textbf{24; 4062} (1981).


\bibitem{102} Chakrabarti, Bikas K and Dutta, Amit and Sen, Parongama,    \textit{Quantum Ising phases and transitions in transverse Ising models}, Springer Science \& Business Media (1996).


\bibitem{103}  Kao, Ying-Jer and Grest, GARY S and Levin, K and Brooke, J and Rosenbaum, Thomas Felix and Aeppli, G,     \textit{History-dependent phenomena in the transverse Ising ferroglass: The free-energy landscape}, Physical Review B, \textbf{64; 060402} (2001).

\bibitem{104} Suzuki, Sei and Inoue, Jun-ichi and Chakrabarti, Bikas K,   \textit{Quantum Ising phases and transitions in transverse Ising models},  Springer (2012).


\bibitem{105} Mukherjee, Sudip and Rajak, Atanu and Chakrabarti, Bikas K,    \textit{Possible ergodic-nonergodic regions in the quantum Sherrington-Kirkpatrick spin glass model and quantum annealing}, Physical Review E,  \textbf{97; 022146} (2018).


\bibitem{106} Leschke, Hajo and Manai, Chokri and Ruder, Rainer and Warzel, Simone,      \textit{Existence of replica-symmetry breaking in quantum glasses}, Physical review letters,  \textbf{127; 207204} (2021).

\bibitem{107}  Yaacoby, Ran and Schaar, Nathan and Kellerhals, Leon and Raz, Oren and Hermelin, Danny and Pugatch, Rami,  \textit{Comparison between a quantum annealer and a classical approximation algorithm for computing the ground state of an Ising spin glass}, Physical Review E, \textbf{105; 035305} (2022).


 \bibitem{108} Schindler, Paul M. and Guaita, Tommaso and Shi, Tao and Demler, Eugene and Cirac, J. Ignacio,  \textit{A Variational Ansatz for the
Ground State of the Quantum Sherrington-Kirkpatrick
Model},  arXiv preprint 2204.02923, (2022).


\bibitem{109} Apolloni, Bruno and Cesa-Bianchi, Nicol{\`o} and De Falco, Diego,    \textit{}A numerical implementation of  ``quantum annealing'', Stochastic Processes, Physics and Geometry: Proceedings of the Ascona-Locarno Conference,  \textbf{97-111} (1990).


 \bibitem{110} Hauke, Philipp and Katzgraber, Helmut G and Lechner, Wolfgang and Nishimori, Hidetoshi and Oliver, William D,      \textit{Perspectives of quantum annealing: Methods and implementations},  Reports on Progress in Physics, \textbf{83; 054401} (2020).

 
 \bibitem{111}  Starchl, Elias and Ritsch, Helmut, \textit{Unraveling the
origin of higher success probabilities in quantum
annealing versus semi-classical annealing}, Journal of Physics B: Atomic Molecular \& Optical Physics, \textbf{55; 025501} (2022) \href{https://iopscience.iop.org/article/10.1088/1361-6455/ac489a}{https://iopscience.iop.org/article/10.1088/1361-6455/ac489a}.

 
 \bibitem{112}  Benioff, Paul,  \textit{The computer as a physical system: A microscopic quantum mechanical Hamiltonian model of computers as represented by Turing machines}, Journal of statistical physics,   \textbf{22; 563--591} (1980).


 
 \bibitem{113}  Deutsch, David,   \textit{Quantum theory, the Church--Turing principle and the universal quantum computer}, Proceedings of the Royal Society of London. A. Mathematical and Physical Sciences, \textbf{400; 97--117} (1985).
 
 
 \bibitem{114} Adrian Giordani,  \textit{Nobel Prize goes to Quantum Computing pioneers}, SCIENCENODE (2012), \href{https://sciencenode.org/spotlight/nobel-prize-goes-quantum-computing-pioneers.php} {https://sciencenode.org/spotlight/nobel-prize-goes-quantum-computing-pioneers.php}.  

 \bibitem{115}  Tanaka, Shu and Tamura, Ryo and Chakrabarti, Bikas K,     \textit{Quantum spin glasses, annealing and computation},  Cambridge University Press, Cambridge (2017).

 \bibitem{116}  Kutner, Ryszard and Schinckus, Christophe and Stanley, H Eugene,     
 \textit{Three Risky Decades: A Time for Econophysics?}, Entropy,  \textbf{24; 627} (2022).

 




 
\end{thebibliography}
\end{document}